\documentclass[12pt
]{article}

\usepackage{natbib}
\usepackage{hyphenat}
\usepackage{hyperref}

\newcommand{\tit}[1]{\textit{#1}}
\newcommand{\tbf}[1]{\textbf{#1}}

\begin{document}

\title{French on London and Bauer, and QBism}

    \author{Jacques L. Pienaar \\ {\footnotesize Instituto de Física, Universidade Federal do Rio de Janeiro} \\ \footnotesize{149 Av.\ Athos da Silveira Ramos, Cidade Universitária,} \\ \footnotesize{Rio de Janeiro, RJ 21941-909, Brazil} \\ {\footnotesize jacques.pienaar@zoho.com}}


\maketitle

\newpage
\section*{Abstract}
In this article I compare two interpretations of quantum mechanics (QM) that draw inspiration from phenomenology: the London-Bauer-French interpretation (hereafter LBF) as articulated by Steven French~\citep{LBF}, and QBism~\citep{QBism_perimeter,QBism_handbook,QBism_Mermin}. I give special attention to certain disagreements between QBism and LBF identified in~\cite{french_flesh, LBF}, as well as French's related claims that QBism may be at odds with key ideas in phenomenology. My main finding is that QBism does not fare so badly with phenomenology as French makes out; in particular it \tit{can be} made compatible with Zahavi's \tit{correlationism} and Husserl's notion of intersubjectivity, both of which strongly inform LBF. Nevertheless, I concur with French's argument that QBism is incompatible with the conception of quantum measurement in LBF, hence also with that of Merleau-Ponty, as the latter based his own analysis on that of London \& Bauer. I explain why I find QBism's account preferable in this case.

\section{Introduction}
In ``A Phenomenological Approach to Quantum Mechanics: Cutting the Chain of Correlations"~\citep{LBF}, French revitalizes and elaborates upon the classic manuscript by London \& Bauer~\citep{LB_engtrans,LB_1939} dealing with the quantum measurement problem, clarifying debates surrounding the meaning of this work by situating it firmly in the phenomenological tradition. The resulting interpretation is hereafter named ``LBF'', which stands for ``the London-Bauer-French interpretation''. London \& Bauer's work may therefore be considered the earliest serious attempt at making sense of quantum theory using the tools of phenomenology; the question may then be raised as to whether it is the \tit{only} viable phenomenological approach to understanding quantum mechanics. 

More recently, a body of work has revealed a number of affinities between phenomenology and QBism (formerly ``Quantum Bayesianism''\footnote{As \cite{stacey_enroute} explains, earlier articles that are often cited as QBist, such as that by \cite{CFS_2002}, are actually precursors to QBism and do not accurately represent its views. QBism was only consolidated as a relatively stable interpretation years later, with the term ``QBism'' only appearing in the literature from 2009 onwards.}), an interpretation of quantum mechanics articulated by Fuchs \& Schack from approximately 2009 (with several changes and refinements since then)~\citep{FS_coherence,FS_greeks,QBism_Mermin,QBism_perimeter,fuchs_PR,fuchs_nwb,FAQBism,QBism_handbook}. QBism explicates the meaning of the quantum formalism in terms of the experiences and expectations of a situated, engaged subject -- an \tit{agent}. This makes it a natural target for phenomenological exegesis which has inspired a number of recent articles (for a sample, see the articles collected in~\cite{phenomenology_QBism_volume,wiltsche_phenomenological_2020}). A phenomenological reading of QBism would therefore provide a useful foil against which to compare LBF.

In order to compare these two approaches, one must first address an issue that arises due to their distinctive histories. In formulating his version of London and Bauer's interpretation, French pointed out that there is good evidence London brought his own philosophy, which was rooted in the phenomenology of Husserl, to bear upon the quantum measurement problem in his manuscript with Bauer. As such, LBF was infused with ideas from phenomenology from the beginning.

QBism's history is much more eclectic in terms of its philosophical influences. As originally conceived, QBism was not inspired by philosophy at all, but grew out of a productive union between the formal apparatus of quantum information theory and a broadly Bayesian interpretation of probabilities. Its only ``philosophical" component in those early days was its commitment to a particular brand of \tit{personalist Bayesianism} associated with Ramsey, Savage, and de Finetti, and of the three only Ramsey was a philosopher \tit{stricto sensu}.

Much later, by chance, Fuchs discovered the work of William James, and through him the American Pragmatist movement which became (and remains) QBism's principal guiding philosophy. Since then, QBism has found common ground with a bewilderingly diverse range of philosophical schools of thought, including \tit{Enactivism}~\citep{varela_embodied_1992}, Barad's \tit{agential realism}~\citep{barad}, Whitehead's \tit{process philosophy}~\citep{Whitehead}, and specific philosophers including Jeffrey, Hacking, and van Fraassen, to name a few; phenomenology is but a recent addition to the list.

This is not to diminish the relevance and impact that phenomenology promises to have on the shape of QBism; however the fact that QBism was not formulated with phenomenology in mind makes it unsurprising that there are some points on which the two do not quite align, and some new work may be needed on one side or the other to eliminate the friction. 

Another consequence of their different histories is that, while LBF is by its nature a phenomenological interpretation of QM, there is currently no single authoritative phenomenological reading of QBism; there are only some proposals, some more complete than others -- a ``phenomenological QBism", to the extent that it exists at all, is still a work in progress. In order to compare the two interpretations, therefore, it is especially important to tease apart those points of disagreement which stem from QBism's being in tension with LBF \tit{specifically}, from those which stem from QBism's being in tension with phenomenology \tit{generally}. This is one of the main aims of the present work. My main finding is that the points of tension that French identifies as being of the latter kind, are in fact chiefly of the former kind. 

In Sec.~\ref{sec:LBF} I review LBF, focusing on the elements central to my arguments. In Sec.~\ref{sec:versus} I briefly introduce the three key points of tension between LBF and QBism that will be the focus of the rest of the paper. Sec.~\ref{sec:introspection} deals with `introspection', which leads naturally to the topic of Sec.~\ref{sec:split}, in which I criticize what I consider to be a hidden `Cartesian split' within LBF. Sec.~\ref{sec:prolongation} covers the conflict between QBism's `prolongation thesis' on the one hand and Merleau-Ponty's account of quantum measurement apparatuses (and that of LBF) on the other. Finally, Sec.~\ref{sec:intersubjectivity} compares the two on the themes of intersubjectivity and agreement of measurement outcomes. The Conclusion follows.

\section{The London-Bauer-French Interpretation \label{sec:LBF}}
In this section I summarize the key elements of LBF that will be relevant for this article. The key to LBF is to provide a phenomenological account of London \& Bauer's `little book'~\citep{LB_1939, LB_engtrans}, which focuses on the act of measurement in quantum mechanics: an observer measures some property of a quantum system using a quantum measurement apparatus. As French explains, London \& Bauer's explication of quantum measurement was widely misunderstood by physicists, who interpreted it as endorsing a ``consciousness causes collapse" interpretation. This was due in large part to confusion about what the authors meant by the observer's ``characteristic faculty of introspection", and in particular how it could be reconciled with their simultaneous claim that the joint state of observer plus system was entangled. French argues -- convincingly, I think -- that the confusion is eliminated when one takes into account London's background in phenomenology and accordingly interprets these claims by London \& Bauer through a phenomenological lens; the result is LBF.

For present purposes, it is enough to focus on the following three core principles upon which LBF is built:\\

\noindent (1) A commitment to a \tit{correlationist} view of phenomenology articulated by Zahavi~\citep{zahavi_2017}, which posits the `correlation' as a fundamental structure, antecedent to (and implicit in) the subject-object split; \\ 
\noindent (2) The postulate that the correlation involving the observer and system after the measurement is objectively ``represented" or ``described" by the entangled wavefunction;\\
\noindent (3) The postulate that the ``characteristic act of introspection" which according to London \& Bauer ``breaks the chain of correlations" is to be interpreted phenomenologically (following Gurwitsch~\citep{gurwitsch1941}) as a ``reflective act" that gives rise to the subject and object ``poles" of the correlation, these being associated with the observer's ``ego" and the measurement result's ``objectification", respectively.\\

These principles are not exhaustive; I single them out here because they are the nexus of the disagreement between LBF and QBism, as I explain in the next section. Before proceeding to that topic, however, it is necessary to make a few pertinent remarks about these three principles that will clarify their meaning and prepare the ground for my subsequent argumentation.

Firstly, a word about Zahavi's \tit{correlationism} mentioned in principle (1). \cite{LBF} uses ``correlationism" somewhat loosely, to encompass not only the phenomenological view articulated by Zahavi, but also French's own suggestion in (2) that the correlations are to be associated with quantum entangled states. This may give the careless reader the impression that this is the only one valid way to interpret Zahavi's `correlation' by the lights of quantum theory. Later in Sec.~\ref{sec:split} I will argue that there is at least one other possibility, which is suggested by QBism; for the moment I simply point out that (2) is an additional postulate connecting correlationism to the quantum formalism, which by no means follows from correlationism as conceived by Zahavi.

Secondly, regarding principle (2), in saying that the correlation is ``represented" or ``described" by an entangled state, French does not imply a ``naive realist" view of the entangled state; that is, he does not endorse a \tit{psi-ontic} interpretation according to which wavefunctions are in one-to-one correspondence with physical reality~\citep{harrigan_einstein_2010}. If that were the case, the correlation described by the wavefunction would be a strictly physical entity, and French is emphatic that the correlation, insofar as it concerns reality, is inseparable from consciousness: it goes beyond any distinction between the strictly mental and physical\footnote{French quotes ~\cite{zahavi_2017}: ``objects have their essentially manifest properties even when not being experienced, and can also truthfully possess them before the emergence of conscious creatures and after their eventual extinction [...] but as essentially manifestable, they do not have a nature that transcends what can be given in experience'', and Husserl's notion of a `transcendental correlation between 
world and world-consciousness', in opposition to ``the `philosophical absolutizing' of the world which is inherent to 
metaphysical realism''~\citep{LBF}(p.100-101). For this reason I have elected to refer to the correlations as `transcendental' rather than `metaphysical'.}. This is perfectly in keeping with Zahavi's conception of correlationism, which in turn follows phenomenology's general rejection of the Cartesian mind-world distinction as fundamental. Indeed, for French, the whole point of invoking correlationism here is to enable the peaceful co-existence of the entangled observer-apparatus-system state with the facts available to the observer's own consciousness, \tit{ie.}\ the perception of a unique measurement outcome. The heart and soul of LBF lies in the explication of how this reconciliation can be achieved without on the one hand endorsing a ``consciousness-causes-collapse" interpretation, nor on the other hand denying consciousness a central role in quantum measurement.

This brings me to my final remark, concerning principle (3). French, following London \& Bauer, is adamant that the entangled conjoined state of observer-apparatus-system is valid from an ``external" point of view, while at the same time the observer in question -- by virtue of their ``characteristic faculty of introspection" -- is ``entitled", or ``has the right" to assign a quantum state corresponding to the outcome they observe. To avoid conflict, French emphasizes that the latter point of view is ``internal", as it were, to the observer, and has a phenomenological, rather than physical, significance (for otherwise it would be construed as being in contradiction with the ``external" state assignment). 

French's language surrounding this delicate issue is extremely circumspect: he often places ``internal" and ``external" in scare-quotes, as I do, and this is eminently wise, for -- borrowing the words from Zahavi -- ``The very division between inner–outer is precisely one with which phenomenology plays havoc"~\citep{Zahavi_basics}. But if French is not suggesting by this language that we are comparing views that are literally internal or external to the observer's mind, then what does he mean by these terms? I will discuss this later on in Sec.~\ref{sec:split}, with the aim of showing that French's internal-external distinction actually conceals an implicit subject-object distinction that undermines his claim to have escaped Cartesian dualism. For now, it is sufficient to highlight that in order to reconcile the ``internal" quantum state assigned by the observer with the ``external" entangled state, it is necessary to regard these state assignments as associated with distinct ``perspectives" on one and the same transcendental correlation. 

\section{QBism versus LBF \label{sec:versus}}
My focus in this article is on the fact that London \& Bauer's account of quantum measurement, and hence that of LBF, disagrees with QBism's account on several important points. French has already identified and commented on this discrepancy~\citep{LBF,french_flesh}, and I take these two references as my main source material. What are the roots of this disagreement, and what does it signify for phenomenological approaches to quantum mechanics?

Given that LBF is, as I have noted, steeped in phenomenology to its very roots, while QBism's relation to phenomenology is still at the stage of dipping in its toes (so to speak), it is natural to wonder whether QBism's disagreements with LBF stem from an underlying disagreement between QBism and certain ideas of phenomenology itself. French certainly thinks so, and though he does not go as far as denying that QBism admits \tit{some} phenomenological interpretation, he has argued that QBism's tenets are in tension with correlationism~\citep{LBF}, and has questioned its ability to account for \tit{intersubjectivity}, a concept defined by Husserl that is very important to phenomenologists~\citep{french_flesh}. In making his arguments, French frequently contrasts QBism's position to that of LBF, in support of which he cites key ideas and passages from the work of prominent phenomenologists Zahavi, Gurwitsch, Merleau-Ponty, and of course Husserl.

In the following sections I critically re-examine French's main arguments. Specifically, I will address the following claims found in~\cite{LBF,french_flesh}: \\

\noindent (A1).\ QBism does not permit the observer to assign a quantum state to themselves, making it incompatible with LBF, in which the observer has a ``characteristic faculty of \tit{introspection}", to be understood following Gurwitsch as a \tit{reflective act} by which the observer ``keeps track of their own state";\\

\noindent (A2).\ QBism asserts that the quantum measurement apparatus is an extension of the observer's senses; this explicitly contradicts the treatment of apparatuses in LBF, and moreover contradicts certain remarks of Merleau-Ponty. French argues that this moreover places QBism in direct tension with Zahavi's \tit{correlationism}; \\

\noindent (A3).\ Related to (A2) above, QBism's view of apparatuses denies it the resources to explain the \tit{intersubjective} character of measurement outcomes, whereas in LBF the apparatus is taken to be external to the observer and macroscopic, which allows it to accommodate intersubjectivity.\\

In the sections that follow, I will examine these arguments in turn.

\section{LBF versus QBism on ``introspection" \label{sec:introspection}} 
Recall from Sec.~\ref{sec:LBF} that, in order to solve the quantum ``measurement problem", in LBF the act of quantum measurement is grounded in two complementary and mutually applicable concepts: first there is the \tit{correlation}, a basic relation that is constitutive of (rather than supervenient on) its own relata, which exists conceptually prior to the observer-object distinction; second there is the \tit{reflective act}, whereby both an observer and observed outcome are implicated ``within" the correlation. 

The correlation and the reflective act are concepts taken from phenomenology; in order to make the connection to quantum measurement, additional assumptions must be made. To this end, LBF contains the following two postulates: (i) the correlation is represented ``externally" (as it were) by an entangled quantum wavefunction, and (ii) the so-called ``collapse of the wavefunction", which singles out a unique outcome from among the possibilities, is not a physical process in the usual sense but rather a reflective act in the phenomenological sense, which London \& Bauer call ``introspection". 

Let us contrast this with the way QBism regards quantum measurement. In QBism, quantum states, whether collapsed or not, never \tit{describe} or \tit{represent} of states of affairs; instead they are equivalent to catalogs of subjective probability assignments that quantify an agent's expectations about what they might experience in response to various actions they could take in the world. That is, in QBism a quantum state quantifies an agent's \tit{degrees of belief} about the outcomes they might obtain for any quantum measurement they could perform. Importantly, measurement outcomes are conceived of as \tit{experiences} the agent could have, hence are \tit{personal} in that they are contingent on that particular agent's embodied possibilities of perception. Also, in keeping with the subjective Bayesian approach to probability theory, there is no extrinsic criteria by which to ascertain that a given agent's probability assignment (hence state assignment) is right or wrong in some absolute sense -- all that can be ascertained is whether the agent's probability assignments are mutually \tit{coherent} among themselves. 

For present purposes, two important corollaries of the QBist view are that (i) the agent cannot meaningfully assign a quantum state to themselves, and (ii) there can be no concept of a quantum state whose existence is independent of an agent who assigns it. I will now elaborate on these points.\\

\noindent \tbf{(i).} There is a danger of misunderstanding this point. Specifically, QBism requires that, in the moment of deliberation in which an agent assigns probabilities, there is an active component of the agent's awareness which encompasses both the agent's deliberative thought process and their momentary readiness for action. It is these \tit{momentarily present and active} aspects of the agent which cannot be included within the probability assignment, for the strictly \tit{logical} reason that they provide the very conditions of possibility for the assignment to be made, and so cannot themselves be taken as the \tit{objects} of that assignment. In QBism nothing prevents an agent from positing a \tit{model} of themselves in order to predict their own actions and behaviours under some hypothetical circumstances removed from their present activity, however, the ``self" which is thereby modeled \tit{is not the self that is doing the modeling}. Attempting to directly include the \tit{model-making self} within the very model that it is making would lead inevitably to logical self-reference paradoxes, making it internally inconsistent.\\

\noindent \tbf{(ii)}. The idea that there can be no state without an agent to assign it follows readily enough from the principle that quantum states are personal probability assignments. However, this does not imply that quantum states cannot have a \tit{de facto} objective character, if `objective' is understood as arising through a process of inter-subjective co-ordination; I return to this point in Sec.~\ref{sec:intersubjectivity}. A good example of this are the articles published in physics journals, where it is commonplace to find statements asserting that, in a given proposed laboratory set-up, the quantum state \tit{is} such-and-such. That such statements are accepted as valid, despite the state assignment apparently not implicating any particular agent, is not evidence against the QBist doctrine; the point is rather that one must always regard such `objective' assignments as arising from a process of \tit{objectification} that involves the active participation of many agents, and supervenes on the ultimately subjective assignments of each individual.\\

Now, French takes the no-self-state-assignment principle (i) to be intrinsically at odds with phenomenology, because (he supposes) the former implies that the observer cannot have access to their own ``mental state" from moment to moment, and therefore excludes the phenomenological concept of introspection. 

However, when London \& Bauer define introspection as the observer's ability to ``keep track from moment to moment of his own state", the introduction of the word ``state" would be \tit{prima facie} puzzling, if ``state" were taken to mean something akin to the \tit{physical state} of an object, \tit{viz.}\ a specification of all of its properties at a given moment. On the contrary, French himself avers that introspection \tit{qua} reflective act is:
\begin{quote}
     [\dots] a `characteristic' act that we perform all the time, from moment to moment, as we observe the world around us. Normally we do not explicitly ‘keep track’ of our mental states, in the sense of making a note of them, say, but [\dots] we do possess this ‘characteristic faculty’ and can say what our state is, if needs be. --- \citep[p.127]{LBF}
\end{quote}

None of this requires the observer to assign a state to themselves in the sense understood by physicists, which even in the simplest case would entail the explicit construction of a mathematical model, in this case requiring the observer to model their own mind. 
It follows that the apparent conflict is illusory: QBism can consistently maintain that the agent is capable of introspection in the sense meant by London \& Bauer, without thereby assigning a physical state to themselves\footnote{It is worth mentioning that the principle (i) is common to both QBism and to a rival interpretation, relational quantum mechanics (RQM)~\citep{Pienaar2021_QBRQM}; however the justification for it in each case is quite different. Unlike QBism, RQM treats observers as purely physical systems, and so probability assignments are not attributed to any conscious agency doing the assigning, hence RQM cannot appeal to the same argument that QBism uses. Instead, RQM argues that the observer is akin to a material `reference frame' relative to which the target system is defined, and claims that therefore the frame ``cannot have information about itself" because such information would presuppose another extraneous frame against which it's own state could be compared. In~\cite{LBF} French criticizes RQM's justification for the no-self-state-assignment principle, but his criticism there does not touch the justification that QBism gives for this principle, which I have explained above.}.
That being said, leaving aside introspection, there is another aspect of LBF that \tit{does} require the observer to assign a quantum state to themselves, and this \tit{is} in conflict with QBism, as I discuss in the next section.

\section{The concealed Cartesian split of LBF \label{sec:split}}
In his critique of QBism~\citep{french_flesh}, French remarks that
\begin{quote}
    The	crucial	difference between London and Bauer’s account, that Merleau-Ponty drew on, and the QBists is that the former, unlike the latter, explicitly take the observer to be entangled with the apparatus and the system. --- \citep{french_flesh}
\end{quote}
In a footnote, he adds: ``Indeed, they accept what Fuchs denies, namely that consciousness can enter into a superposition". In light of our exposition of the QBist no-self-state-assignment principle in the previous section, by ``consciousness'' French presumably means to refer to the very consciousness that is actively assigning the state; for as we have seen, this is what is relevant to QBism's ban on superpositions of consciousness\footnote{Indeed, French's careful critiques of QBism suggest he is well aware that QBists \tit{do} allow an agent to assign a quantum superposition state to \tit{another} agent, without thereby denying that the other agent is a conscious being.}. The point being made here, then, is that LBF is not so constrained as QBism in this regard, because it takes the entangled state of observer-apparatus-system as playing a descriptive role, which QBism denies. I raise no objection to this assessment; it appears to me correct.

However, French goes one step further when he suggests that this prevents QBism from being compatible with Zahavi's correlationism. I find this further claim unjustified: for as I remarked in Sec.~\ref{sec:LBF}, the postulate in LBF that links the correlation to an entangled state is just \tit{one} way of relating the former concept to the quantum formalism; there may be others. 

Indeed, QBism offers at least two possible alternatives: taking a structural realist approach, QBism's form of the Born rule (the so-called \tit{Urgleichung}) fits the bill: it represents a fundamental and universally valid constraint on the \tit{relation} between the agent and their external world. Alternatively, one could focus on the measurement outcomes themselves, taking these to represent something akin to a primordial experience conceptually prior to the differentiation of subject and object; an approach along these lines is articulated in~\cite{pienaar_2020}.

What, then, is the true point of conflict between QBism and LBF? It is that in LBF, the analysis of the measurement only ever involves a \tit{single} observer, yet it refers to \tit{two} alternative state assignments that are supposed to both be valid for that observer: the entangled state of observer-apparatus-system, which is valid from the ``external" view, and the collapsed state which is valid from the ``internal" view. Contrast this with the QBist position in which the only state that is `valid' for an observer is (by definition) the one they themselves assign; in a catchphrase, \tit{one observer, one state}.  
The problem with French's internal-external distinction is that, when one examines the role that this duality plays within LBF, one finds that it \tit{functionally} performs the role of an object-subject split, that is, it represents the Cartesian split in disguise. The following passage is revealing (my emphasis): 
\begin{quote}
    As we have seen, the observer is included in the superposition - at least from the external perspective. Internally, as it were, the observer does become separated and in that separation, is no longer described by the formalism [\dots] From this internal perspective, the observer, as an ‘I’ or ego, \tit{is not a ‘natural’ object at all, but rather a phenomenological one}. There simply is no possibility of describing the observer in this sense in quantum mechanical or any other physical terms - indeed, there never was.\citep[p.130]{LBF}
\end{quote}
Here, finally pressed to elaborate on the meaning of his external-internal distinction, French aligns it with a distinction between `natural' versus `phenomenological' objects, where only the former are to be described `in physical terms'. This seems to raise a question regarding the physical meaningfulness of the ``collapsed" state assignment that the observer is ``entitled" to make.

To sharpen the dilemma, note that according to French, the ``external" entangled state assignment is supposed to be \tit{correct}, in the sense that it accurately predicts the outcome statistics for any measurements performed on all or part of the observer-apparatus-system conjunction. It follows that the non-entangled state that the observer might assign from their ``internal" point of view would lead to \tit{incorrect} predictions for at least some of these measurements -- that is, it cannot be \tit{stricto senso} a correct state assignment. 

Now the difficulty is manifest: on the one hand, as remarked in Sec.\ref{sec:LBF}, French is at pains to distance LBF from interpretations that take quantum states to be exhaustive descriptions of reality, such as the many-worlds or psi-ontic interpretations. On the other hand, in order to avoid claiming that wavefunction collapse is a causal process (which would be anathema to the spirit of LBF), the ``reflective act" by which the observer apprehends a unique outcome cannot imply any kind of change that could be described `in physical terms'. Between this proverbial Scylla and Charybdis, the status of the ``collapsed" state sits uncomfortably. 

One thing is clear: French cannot simply classify the ``collapsed" state as \tit{invalid} and drop it; that would leave only the external state assignment -- the universal wavefunction -- and then LBF would be essentially the many-worlds interpretation with a phenomenological gloss. Instead, French uses various rhetorical techniques to argue that the ``collapsed" state remains valid in a `phenomenological' sense. The transformation from entangled to collapsed state is not causal, but more akin to a ``shift in perspective" from the `external' to the `internal' point of view. The phrase ``collapsed state", with its overtly physical connotations, is eschewed in favour of ``separated state" or ``state post-separation", referring to the separation of ego from object in the reflective act. Unlike the entangled state, the collapsed state is never referred to as a ``correct" description, but merely as a state that the observer is ``entitled" to assign (the implication being, I suppose, that one is entitled to be sometimes incorrect). And so on. 

The problem with this strategy is that, whatever we say about it, the collapsed state is still a quantum state assignment that implies certain statistical predictions, and on the presumption of the correctness of the entangled state, there exist experiments whose data would show predictions predicated on the collapsed state to be incorrect. Now, this might be avoided if it turned out that the experiments in question were of such a global nature as to be only accessible to the external viewpoint, while remaining unavailable in principle to the observer implicated within the measurement. In that happy circumstance, the observer would never encounter any evidence contradicting their ``collapsed" state assignment, and one could argue that in this way it maintains a sort of relational validity, relative to the given observer.

However, perhaps surprisingly, this is not true, as demonstrated by~\citep{baumann_brukner_2020}, using an ingenious twist on Deutsch's version of the Wigner's friend thought experiment. In this version, Wigner measures the friend in a superposition basis (that is, observes the interference terms of the entangled superposition that includes the friend) and then communicates the outcomes of his measurement to the friend in such a way that the superposition state of the latter is preserved. Given the premise that the entangled state assignment made by Wigner is the ``correct" one -- that is, assuming the statistics of Wigner's measurements bear witness to the entanglement -- the friend is forced to reject their own ``collapsed" state assignment, as it proves inadequate to explain the data they receive from Wigner.

Note that if we reject the basic premise that the outcomes actually observed by Wigner will vindicate his entangled state assignment, then no problem would arise. Indeed, following QBism~\citep{wf_fellow}, we could hold that Wigner and his friend's state assignments do not describe states of affairs -- they do not ``tell nature what to do" -- but merely reflect each party's different beliefs about what will happen. 

LBF is denied this option, however, as it bestows a privileged role upon the entangled state, taking it to be descriptive of the underlying correlation. French's rhetorical maneuvering notwithstanding, there is then no recourse to claiming that the ``collapsed" state retains any validity as to physical states of affairs; consigned to the mind of the observer, it's only role is to signify the existence of a personal, subjective mental world over and above the public, objective physical world that is described by entangled states. The result is that the phenomenological aspect of LBF does not actually create any trouble for those seeking to interpret quantum theory along the lines of many-worlds; it merely dresses the latter with a `phenomenological' layer that floats on top of the universal wavefunction like so much flotsam (as it were), not disrupting its basic metaphysical shape.

\section{LBF versus QBism on the nature of the quantum measurement apparatus \label{sec:prolongation}}
One of QBism's more controversial tenets, to which it nevertheless holds firm, is the claim that the ``quantum measurement apparatus" is an extension of the agent's senses. 

French has already pointed out that QBism's ``prolongation thesis'' is in direct conflict with LBF, and moreover flatly contradicts Merleau-Ponty's position on this matter~\citep{LBF, french_flesh}. In this section, I will not take issue with French's claim, which I think is correct; rather, I will criticize Merleau-Ponty's argument, showing that it fails to articulate the proper conditions for an instrument -- classical or quantum -- to serve as an extension of the senses, and therefore does not pose a serious challenge to QBism's prolongation thesis.

First, some explanation is in order: for, if taken at face value, one might think the prolongation thesis is patently absurd. In most real examples of scientific measurement (quantum or classical) the scientist obviously does \tit{not} directly experience the measurement outcome, but rather they experience the apparatus interface -- its dials and blinking lights, let us say -- making it phenomenologically an \tit{object} external to the agent and not a part of themselves. To make sense of the QBist claim, one actually needs to understand it as a radical proposal for a re-definition of what ``apparatus" should mean~\citep{pienaar_2020}. Insofar as the agent experiences an instrument as an external object, the QBist says the instrument \tit{is not} an apparatus. Rather, in such commonplace scenarios, the true ``apparatus" is comprised of the agent's own bodily senses, and the measured system is not, in fact, the electron (or whatever the instrument is probing) but rather the instrument itself. On the other hand -- and this is the salient point for our present discussion -- QBism says that the determination of what constitutes part of the agent's body versus what counts as an external instrument is not given once and for all, but is in principle completely open; it is decided not by principle but practice, that is, the body-world boundary is enacted dynamically through the process of using the instrument to make a measurement\footnote{In their ongoing quest to better articulate their position with regards to the body-world boundary, QBists take inspiration from the philosophy of embodied cognition (``enactivism")~\citep{varela_embodied_1992}, as well as the ``New Materialists", particularly Karen Barad~\citep{barad}.}. 

The paradigmatic example underlying this idea is using a cane to ``feel" the shape of the ground: to the extent that one grips the cane firmly, and is sufficiently experienced in its use to immediately grasp also the meaning of the vibrations transmitted through it, then it functions as if it were an extension of their body: the user is then not aware of the cane, but rather of the ground, as though they were touching it directly. QBism says, in this case, we can drop the ``as if" language: the cane \tit{is} a part of the user's body, for in order to be part of one's body it is entirely sufficient that it be experienced as a body-part. This tenet of QBism, called the ``prolongation thesis" for brevity, is thoroughly explicated in~\cite{pienaar_2020, Pienaar_entities}, so I will not enter into details here. Suffice to note that, while freely admitting that it is far-fetched, QBists do not exclude that an agent with (say) ``electron microscope eyes" could be said to literally sense electron spins\footnote{\tit{cf}.\ the extended discussion of the `argument from mutation' in \citep{wiltsche_2012,Pienaar_entities}).}. 

With this clarification about the meaning of the prolongation thesis, I now briefly explain why LBF cannot consistently accommodate the thesis, and why it therefore represents a critical point of incompatibility between LBF and QBism.

To begin with, note that in both LBF and QBism, the observer's consciousness is implicated in measurement, and so they both encounter the problem of \tit{intersubjectivity}, first articulated by Husserl. Namely, what is the basis for the epistemological claim, which is so dear to the ``objective" sciences, that the measurement outcome apprehended by \tit{this} observer is accessible to \tit{other} observers as well? Unless one can provide an account of how intersubjectivity comes about, one is open to accusations of solipsism, which is rejected both in QBism and in LBF. 

This is relevant here because the account of intersubjectivity in LBF -- unlike that of QBism -- depends crucially on maintaining a principled distinction between apparatus and the particular observer who uses it. Allowing this line become blurred or ``dynamic" as it is in QBism would seem to make measurement outcomes irretrievably ``personal", blocking the way to intersubjectivity. We shall pick up the theme of intersubjectivity again in the next section; for present purposes, it provides an explanation as to why accepting LBF entails rejecting any suggestion that the \tit{quantum} measurement apparatus could function as an extension of the observer's senses.  

I now turn to the main topic of this section, which is Merleau-Ponty's specific objection to the prolongation thesis, which French points out was directly inspired by London \& Bauer's account of quantum measurement~\citep[p.168]{LBF}. In an oft-quoted passage from the course notes of Merleau-Ponty's lectures on Nature, which he delivered at the Coll{\`e}ge de France in the 1950s, he asserts that while classically ``the apparatus is the prolongation of our senses," in quantum mechanics it cannot be, because ``[t]he apparatus does not present the object to us" but rather ``realizes a sampling of this phenomenon as well as a fixation"~\citep{Merleau-Ponty_nature}. While there is much that is ambiguous in this brief remark, there can be no doubt that French is correct in identifying it as a challenge to QBism's prolongation thesis.

To better understand what Merleau-Ponty means by the above statement, it is useful to break it up into parts. First, there is the claim that classically, the apparatus can be regarded as a prolongation of the senses. Why might this be so? Merleau-Ponty tells us -- implicitly -- it is because it ``presents the object to us". However, this alone cannot justify regarding the apparatus as a ``prolongation of our senses", for the latter requires something more, namely, that the apparatus presents the object to us \tit{in a similar way} to how our native, unaided senses present objects to us; without this proviso, the apparatus would be merely an additional mode of presentation, parallel to the senses, rather than a ``prolongation" of them.

Secondly, Merleau-Ponty claims that quantum mechanically the apparatus does \tit{not} present any object to us. Reading between the lines somewhat, I take it that Merleau-Ponty uses ``sampling" and ``fixation" here in allusion to ``statistical sampling" and the corresponding ``fixation" of a particular value of a random variable. His point is, then, that classical objects can only be presented to us by virtue of the fact that we may take all of their properties (whether directly measured or not) to have determinate values prior to measurement, whereas it is characteristic of quantum measurements that at least some properties are necessarily non-determinate prior to measurement: the theory is ``irreducibly" probabilistic in this sense. Now, insofar as the probabilities for measurements are captured by the wavefunction, the latter is what quantum theory substitutes in place of the classical ``object" -- but of course the wavefunction itself simply can't be regarded as an object, precisely because it never completely specifies determinate values of all properties. So, to close off the argument, quantum measurement apparatuses per definition cannot present us with any object, so cannot be prolongations of our senses.

As straightforward as the argument seems, it relies on some highly questionable premises. 
It presumes throughout that apparatuses aim to measure \tit{objects}, when in fact the whole history of measurement shows that they are designed to measure \tit{properties}. The object, which is supposed to be the bearer of the properties, is of secondary importance in measurement science; its existence may be inferred whenever a set of properties coincide in a stable manner for some length of time, but there are many cases where the object of measurement is unclear, ambiguous, or irrelevant. When measuring atmospheric temperature or pressure at some geographic co-ordinates, what is the measured object? It can't be the whole atmosphere, which has no well-defined value of these properties. When measuring the decibels of noise in some vicinity, what is the object being measured? A sound wave? The instrument producing the sound? More examples in a similar vein can easily be found. 

Now, if we grant that apparatuses measure properties, not objects, the lack of a ``quantum object" need not trouble us in itself. Presumably, then, it is the probabilistic character of quantum properties that sets them apart? However, this too is confounded by actual measurement science. In practice, all measurements of classical quantities only yield probabilities for the values to lie within certain ``confidence intervals" -- the idea that apparatuses present us with determinate properties (let alone objects) is, at least on the face of it, inaccurate.

Similar remarks apply to sense perception: while objects are certainly prominent to our everyday senses, they are by no means necessary for us to have sensory perceptions of qualities. Consider going for a swim in the ocean; certainly a richly sensuous experience, but the sensations, textures, impressions, smells and sounds do not arrange themselves neatly into associations with some set of clearly defined objects. 

That much being granted, even on the question of single properties, there seems to be an important dis-analogy between the classical and quantum case, since the wavefunction generically presents us with superpositions of properties that cannot be thought of as corresponding to unique, definite values prior to the measurement\footnote{More accurately, one \tit{can} think of wavefunctions as `incomplete' descriptions of underlying `hidden' properties, however all such approaches have well-known problems~\citep{sep-bell-theorem}. In any case, Merleau-Ponty's remarks suggest he did not endorse this approach, for instance, he takes the quantum statistics to provide the `maximum image of the object'~\citep[p.93]{Merleau-Ponty_nature}.}. Thus, one may be tempted to say that quantum measurements are \tit{irreducibly} probabilistic, in contrast to the classical case, where (it is usually supposed) measurement uncertainty merely represents ignorance of the underlying true value of the measured property.

Perhaps one could therefore revive Merleau-Ponty's argument by arguing that in order for a measuring instrument to function as a prolongation of the senses, it should present us with a property that has a transcendent existence beyond the measurement act, that is, it should be consistent with the hypothesis that the measurement reveals an independently existing quantity\footnote{It is actually quite difficult to make this case, since perceptions of properties are often `secondary qualities', such as colour, sound, or heat, which seem to be inextricable from the sense organs that `measure' them -- but this difficulty only helps my argument.}. However, here again the history and philosophy of measurement science blocks the would-be argument by showing that the concept of a ``true value'' in classical measurement is not nearly as straightforward and problem-free as it may appear to physicists and philosophers outside of the metrology community.

At the time Merleau-Ponty delivered his lectures, the traditional realist assumption about measurement, \tit{viz.}\ that measurement aims to discover the ``true value" of a specified quantity that exists independently of how we measure it, was still largely uncontested among metrologists; however in the latter half of the 20th century it was fiercely criticized on both philosophical and methodological grounds~\citep{gregis_2023}. In particular, it was argued that the supposed ``true value'', if it exists, was in principle unknowable (since one only ever has access to instrument readings, which can only be compared directly to one another) and thus without operational meaning; moreover, in many applications there appeared to be no \tit{unique} value that could serve in the role of ``true value''. 

A proper review of the issues and their impact on metrology is beyond the scope of this article\footnote{Further details of the `epistemic turn' in metrology and the demise of traditionally realist approaches to classical measurement may be found in \citep{decourtenay_2017,Gregis_2015,gregis_2023,mari_MAS,sep-measurement}.}; suffice to note that it led to significant revisions of the two major international guides for metrologists, the \tit{Guide to the expression of uncertainty in measurement} (GUM)~\citep{JCGM_GUM_2008}, and the \tit{International Vocabulary of Metrology} (VIM)~\citep{VIM_2012}. In particular, an explicitly Bayesian approach to uncertainty was adopted, replacing the traditional approach that conceived of measurement error as the difference of the measured value from the true value. Instead, the GUM takes probability to represent a ``degree of belief'', which has the advantage of permitting a unified account of uncertainty due to systematic and random errors~\citep[E.5.2]{JCGM_GUM_2008}. Correspondingly, while the term ``true value'' is retained in the VIM, its definition is carefully worded and accompanied by extensive qualifications so as to free it from any metaphysical commitments~\citep[2.11]{VIM_2012}.

In light of the foregoing, the reasons Merleau-Ponty offers to support his claim that quantum instruments cannot be extensions of the senses therefore appear to apply equally well to classical instruments; for the idea that classical measuring instruments reveal pre-existing unique properties is no less problematic than the corresponding idea in quantum theory (albeit for different reasons). Should we then conclude that even classical instruments cannot be viewed as extensions of the senses? If one finds that conclusion absurd, then the alternative is to conclude that the aforementioned aspects of quantum and classical measurement are not after all obstacles to viewing either one as an extension of the senses. The door remains open to find necessary and sufficient conditions for the prolongation thesis to hold true in both classical and quantum theory.

If indeed classical apparatuses are able to present properties to us in a manner similar to how our senses present qualities to our perception, then we may regard them as prolongations of our senses. But then the criterion works just as well for quantum measurements -- for when it comes to ascertaining a value of a property (as opposed to presenting a whole object), classical measurements are no less irreducibly probabilistic, nor less dependent on inferences, than quantum measurements. 

One may of course question the extent to which instruments are able to present properties in manner similar to perceived qualities, however this question cuts across the quantum-classical divide, an answer in one domain applying just as well in the other. In conclusion, I do not find in Merleau-Ponty's remarks any compelling objection to the prolongation thesis.

\section{LBF versus QBism on ``intersubjectivity" \label{sec:intersubjectivity}}

I now discuss French's contention that LBF possesses, and QBism lacks, the resources necessary to accommodate intersubjectivity and thereby escape solipsism. Briefly, intersubjectivity is based on the phenomenological concept of the \tit{intentional horizon} of a perceived object, which comprises certain aspects (``profiles") of the object that are not directly sensed, but are nevertheless co-given in perception; an example is how, when perceiving a book, one is aware of its having a back side, even when the latter is not in view. According to Zahavi and Overgaard's reading of Husserl (upon which LBF relies), the horizon, with its co-given ``hidden profiles", actually implicates a hypothetical community of \tit{other} observers -- that is to say, observers for whom, if they were present, the hidden profiles would be visible, as it were~\citep{zahavi_inter}.   
 
As mentioned in the previous section, in order to establish that intersubjectivity is possible, in LBF the apparatus is taken to be a physical object necessarily distinct from the observer. Moreover, the apparatus is ``classical" in a very precise sense; in London \& Bauer's words (translated), ``one always has the right to neglect the effect on the apparatus of the `scrutiny' of the observer". French is careful to note that merely being \tit{macroscopic} is not what makes the apparatus `classical' in the sense meant by London \& Bauer; rather, the classicality of the apparatus is reflected by the fact that `scrutiny', according to French, refers to the `reflective \tit{regard}'. Moreover, ``the regard or `scrutiny' does not change or affect the apparatus, as an object, in any way, and so a `collective scientific perception' can be created in which a second observer, looking at the same apparatus, will make the same observations"~\citep[p.134]{LBF}. 

QBism's prolongation thesis, as I have noted, bars it from introducing intersubjectivity by this route, so it is necessary here to briefly recap why QBism adheres to this doctrine. The reason is that, in order to give a consistent account of the Wigner's friend thought experiment that does not invoke `hidden variables', nor takes the quantum state as descriptive of states of affairs (as in, say, many-worlds interpretations), it is necessary to allow that the friend's measurement outcome occurs \tit{only for the friend} and not for Wigner. That is, quantum theory as read through QBism forces upon us the idea that, by default, measurement outcomes are to be regarded as \tit{private}, that is, \tit{ab initio} they exist only to the one who measures them. It is worth noting that this conclusion has been recently formalized in a series of `extended Wigner's friend' no-go theorems (see~\cite{schmid_review_2023} and references therein). But once we allow that outcomes are personal to the agent who measures them, it is merely a corollary that the instrument by which the outcome is produced must also be personal to the agent, hence the prolongation thesis.

By the same token, the insistence in LBF that the apparatus be classical puts it potentially in conflict with the aforementioned no-go theorems. For if the friend's observed outcome is not just personal to them, but rather exists in principle for all observers who might choose to witness it (even if they don't actually do so), accepting LBF appears to entail accepting the assumption that~\citep{bong_2020,schmid_review_2023} call `Absoluteness of Observed Events', which is shown to be in tension with other principles of LBF, namely the universal applicability of quantum theory and the rejection of hidden variables. Alternatively, one could try to argue that in LBF the friend's outcome does not `exist' in the sense relevant to the aforementioned theorems, but then the outcome's \tit{non-existence} would be an intersubjectively verifiable physical fact, essentially eliminating the friend's point of view from the realm of physics. One might try to save the legitimacy of the friend's point of view in LBF by enshrining the friend's outcome in the `phenomenological' world, but this seems to strip it of any physical trappings, making it hard to see how anyone else could possibly access it. How are other observers supposed to check the outcome except by perceiving the \tit{physical} pointer needle (say)? The result seems not far different from solipsism after all.

What, then, are QBism's prospects for giving an account of intersubjectivity? French argues they are slim: for if the only way to include the observer in a quantum state is to have that state be assigned by \tit{another} observer, it seems that one can never transcend the observer's point of view, and so can never arrive at an `external' description of the observer-object relation. True enough, but is such an `external' description really necessary to escape solipsism? QBism counters that the observer-object relations can be -- indeed \tit{must} be -- navigated and charted from within the standpoints of particular observers; there simply is no `outside' standpoint. This is not solipsism, provided that the observers are able to share at least part of the world they jointly inhabit.

In service of this idea, it is important to stress that accepting the existence of `truly private' quantum events does not exclude that measurement outcomes can be shared; it only means that one needs to explain how the outcomes can be shared under the right conditions. As Fuchs puts it (my emphasis): ``What we learn from Wigner and his friend is that we all have truly private worlds \tit{in addition to our public worlds}"~\citep{fuchs_interview}. For instance, if Wigner were to measure the same property of the system that the friend measures, it would not contradict QBist doctrine to say that the outcomes of their respective measurements refer back to a single shared element of reality.

The preceding considerations show that, contra French, QBism's prolongation thesis \tit{not} represent a fundamental obstacle to accommodating intersubjectivity. Furthermore, in view of the Extended Wigner's Friend no-go theorems, QBism appears to have a considerable advantage over LBF when it comes to the problem of giving an account of intersubjective agreement consistent with quantum theory.

\section{Conclusions}

In this article I have examined LBF as a phenomenological approach to quantum theory and in particular to resolving the measurement problem, by comparing it with QBism. Although not strictly a phenomenological interpretation, QBism is sufficiently compatible with phenomenological ideas so as to make it a useful reference for comparison. I focused on three putative points of disagreement between LBF and QBism, namely: London \& Bauer's notion of \tit{introspection} (Sec.~\ref{sec:introspection}), the nature of the apparatus (Sec.~\ref{sec:prolongation}), and \tit{intersubjectivity} (Sec.~\ref{sec:intersubjectivity}). 

While I found that \tit{introspection} presents no difficulty in itself, it led to the identification of a concealed point of incompatibility between LBF and QBism, which rests upon the former's taking entangled quantum states as descriptive of the observer-object relation. I took this as the basis for a critique of LBF in Sec.~\ref{sec:split}, arguing that it implies a \tit{de facto} reinstatement of Cartesian dualism. I pointed out that it is also the real culprit behind QBism's apparent incompatibility with Zahavi's correlationism; if one does not identify correlations with quantum states as in LBF, then one has alternative avenues for making QBism compatible with correlationism. 

Regarding the apparatus, I agreed with French that there is a real incompatibility due to the rejection in LBF (maintained by Merleau-Ponty) of what I called the prolongation thesis of QBism: that the apparatus is an extension of the agent's senses. However, I argued that Merleau-Ponty's argument against taking the apparatus as an extension of the observer's senses is fundamentally flawed, focusing as it does on objects rather than properties, and implicitly presuming that classical instruments present property values (if not object) that are already there, as it were, independently of measurement -- a claim that I showed is not supported by contemporary literature in the philosophy of metrology.

Finally, on to the closely related topic of intersubjectivity, I argued that the prolongation thesis -- which French argues blocks QBism from intersubjectivity -- is actually \tit{indispensable} for responding to recent `extended Wigner's friend' no-go theorems, and that by rejecting it LBF is deprived of a satisfactory response to those theorems. Against French's claim that endorsing the prolongation thesis threatens to make QBism solipsistic, I argued that his claim is based on two misconceptions: first, that to avoid solipsism it is necessary to adopt an `external' standpoint on the observer-object relation; second, that intersubjectivity encompasses the requirement that different observers agree on the measurement outcome. To make my point, I drew upon the philosophy of metrology to outline a possible route to an account of intersubjectivity in QBism that would have neither of these features.

\tit{Acknowledgments:} This publication was made possible through the support of Grant 62424 from the John Templeton Foundation. The opinions expressed in this publication are those of the authors and do not necessarily reflect the views of the John Templeton Foundation.


\end{document}